\documentclass[prl,superscriptaddress,showpacs,twocolumn]{revtex4-1}
\usepackage{amssymb}
\usepackage{graphicx}
\usepackage{dcolumn}
\usepackage{bm}
\usepackage{amsmath}

\begin{document}

\title{Type-I and type-II topological nodal superconductors with $s$-wave interaction}
\author{Beibing Huang}
\affiliation{Department of Physics, Yancheng Institute of Technology, Yancheng, 224051, China}
\author{Xiaosen Yang}
\affiliation{Department of physics, Jiangsu University, Zhenjiang, 212013, China}
\author{Ning Xu}
\affiliation{Department of Physics, Yancheng Institute of Technology, Yancheng, 224051, China}
\author{Ming Gong}
\email{gongm@ustc.edu.cn}
\affiliation{Key Lab of Quantum Information, CAS, University of Science and Technology of China, Hefei, 230026, P.R. China}
\affiliation{Synergetic Innovation Center of Quantum Information and Quantum Physics, University of Science and Technology of China, Hefei, 230026, P.R. China}

\date{\today}

\begin{abstract}
Topological nodal superconductors are generally realized based on unconventional pairings. In this work, we propose a minimal model
to realize these topological nodal phases with only $s$-wave interaction. In our model the linear and quadratic spin-orbit couplings
along the two directions break the isotropy in momentum space and introduce effective unconventional pairings on the
Fermi surface. This model may support different nodal superconducting phases characterized by either winding number in BDI class or Pfaffian in
D class at the particle-hole invariant axes. In the vicinity of the nodal points the effective Hamiltonian can be described by either type-I or type-II Dirac
equation; and the crossover between these two nodal points can be driven by external Zeeman fields.
We show that these nodal phases are robust against weak disorders, thus are possible to be realized in experiments with real materials. The smoking-gun
evidences to verify these phases based on scanning tunneling spectroscopy are also briefly discussed.
\end{abstract}


\maketitle

Topological phases are featured by nontrivial topological integer
numbers in their bulk and the associated topological protected
gapless edge states along the boundaries and in the defects
\cite{APS, kitaev}. In the gapped topological superconductors (TSCs),
these localized states can realize the long sought Majorana zero
modes\cite{majorana}, which are anyons satisfying non-Abelian statistics.
These zero energy modes are the basic building blocks for fault-tolerant topological
quantum computation\cite{been, alicea, nayak}, thus have attracted great
attention in theories\cite{fuliang, sau, potter, oreg, roman1, roman2, perge, pientka, stano, brau,
franz, nakosai, coldatom1, coldatom2, coldatom3} and
experiments\cite{mourik, rokhin, das, deng, albre, yazdani, jinfeng, jinfeng2}, in both
condensed matter and ultracold atom physics in the past several years.

Nevertheless, natural materials are scarcely to be TSCs except several exceptions, such as
superconducting topological insulators X$_x$Bi$_2$Se$_3$, where
X = Cu\cite{fuliangberg,Hashimoto}, Sr\cite{guan}, and Ti\cite{Zhiwei, Trang}, and
the spin-triplet superconductor Sr$_2$RuO$_4$\cite{rice, Matzdorf, Luke}. Unfortunately, none
of them have been conclusively identified to be TSCs.  Alternatively, many researches about TSCs are focused on combined systems,
in which the superconducting pairings are introduced
to the spin-orbit coupled systems via the Andreev reflection
mechanism. In this respect the most eminent examples are the surfaces
of three-dimensional topological insulators\cite{jinfeng, jinfeng2}, the spin-orbit coupled
semiconductor nanowires\cite{sau, potter, oreg, roman1, roman2, mourik, rokhin, das, deng, albre} and magnetic
systems with inhomogeneous magnetic texture\cite{perge, pientka,
stano, brau, franz, nakosai} in proximity to s-wave superconductors.
The recent experiments \cite{mourik, rokhin, das, deng, albre, yazdani, jinfeng, jinfeng2}
have observed zero-bias peaks in tunneling spectroscopy, which provide promising evidences for the
Majorana zero modes. These systems still belong to the $p$-wave category\cite{Ivanov, ReadGreen} due to the effective $p$-wave pairing on
the Fermi surface in the dressed basis; they are fully gapped except at the phase boundaries.

Gapless TSCs, which are also called as nodal superconductors, are feasible and have been explored in refs. \cite{wenyu, shusa,
ktlaw1, ktlaw2, youjianbing, tanaka, sato, Kashiwaya}. In the fully gapped TSCs, the zero modes  in the middle gap
are protected by topological invariant defined in the whole Brillouin zone. However, in the nodal TSCs, the topological invariants
can only be locally defined in the  Brillouin zone, and the nontrivial topological invariants can lead to
dispersionless zero-energy flat band along the boundaries. These flat bands have been discussed in many literatures\cite{wenyu, shusa, ktlaw1,
ktlaw2, youjianbing, tanaka, sato, Kashiwaya}, which generally require the exotic pairings, such as $d_{xy}$-wave \cite{sato, Kashiwaya},
$p_x$ pairings \cite{sato, Kashiwaya} as well as other unconventional pairings in non-centrosymmetric superconductors\cite{ktlaw1, ktlaw2, youjianbing, tanaka, sato};
see ref. \onlinecite{RevSchnyder} for more possible nodal superconductors.

Engineering the single particle band structures, in some cases, is much easier than direct engineering the pairings, especially for those unconventional pairings whose
underlying pairing mechanisms are complex and unclear. In this work we propose a minimal model to realize these gapless TSCs with only isotropic $s$-wave interaction. To account for the
asymmetric nodal points in the momentum space, the single particle term consists of a linear spin-orbit coupling (SOC) along one direction and a quadratic SOC along the other
direction. Different types of nodal phases with effective type-I and type-II Dirac equation nodal TSCs can be realized; and the crossover between these two nodal points
can be driven by Zeeman fields. These nodal points are robust against weak disorders, which only slightly renormalizes the momentum-independent parameters from Born
approximation. These topological phases may be realized using the semi-Dirac materials in proximity to a $s$-wave superconductor. The
experimental smoking-gun evidences for these nodal phases are also briefly discussed.

We start from the following model in a square lattice,
\begin{eqnarray}
    H_0=\sum_k c_{\textbf{k} s}^{\dag}[d^x_{\textbf{k}}\sigma_x +d^y_{\textbf{k}} \sigma_y +d^z_{\textbf{k}}\sigma_z-\mu\sigma_0]_{ss'}c_{\textbf{k}s'},
    \label{eq-H0}
\end{eqnarray}
where $c_{\textbf{k}s}$ is the annihilation fermion operator with momentum $\textbf{k} = (k_x, k_y)$ and spin $s =
\uparrow, \downarrow$, $\sigma_{x,y,z,0}$ are Pauli matrices and $\mu$ is the chemical potential. We first focused on
(lattice constant $a =1$),
\begin{equation}
        d^x_{\textbf{k}}=\alpha \sin{k_x}, d^y_{\textbf{k}}=\beta(1-\cos{k_y}), d^z_{\textbf{k}}= \gamma (1-\cos{k_x}).
\end{equation}
Notice that a nonzero $\gamma$ is used to open a gap at ${\bf k} = (\pi, 0)$, thus we have a semi-Dirac dispersion
near ${\bf k} = (0, 0)$. In this model, the first term
is the linear SOC along the $k_x$ direction, while along $k_y$
direction, a quadratic SOC is required, which is the major
difference between our idea and the proposal discussed in previous
literatures for the realization of gapped TSCs and associated Majorana
zero modes. This quadratic dispersion can be regarded as a
consequence of fusing of two Dirac points with opposite winding
numbers along the $k_y$ direction\cite{dane, saha, place}, in which the linear dispersion
along this direction is exactly canceled. Thus the above single
particle model with linear dispersion along $k_x$ direction and
quadratic dispersion along $k_y$ direction can be relevant to
semi-Dirac materials, such as BEDT-TTF$_2$I$_3$ salt \cite{kata},
TiO$_2$/V$_2$O$_3$ multi-layer structure \cite{pardo}, the
anisotropic hexagonal lattices in presence of magnetic field
\cite{dietl}. This model may also be realized in ultracold atoms, in
which the linear SOC can be realized by Raman coupling\cite{Spielman} while the
quadratic SOC can be realized using the experimental approach in
Ref. \onlinecite{wu2016}.

\begin{figure}
    \includegraphics[width=0.35\textwidth]{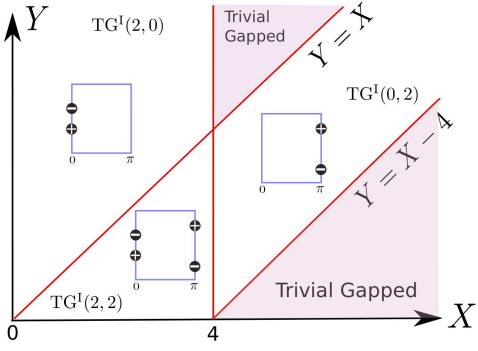}
    \caption{Phase diagram for model (\ref{eq-h2}). The line $X=4$ represents the phase transition takes place at $k_x =\pi$,
    and the line $Y=X-4$ and $Y=X$ represent the transition at $k_x = 0$. The shadowed regimes mark the topological trivial phase TG$^\text{I}(0, 0)$.
    In the nodal TSCs, the symbols $\pm$ in the black circles denote the winding number of the nodal points.}
    \label{fig-fig1}
\end{figure}

Now we introduce Cooper pairs into this system by proximity to a $s$-wave superconductor (in ultracold atoms this
pairing can be realized by attractive interaction). Then we have the following equation,
\begin{eqnarray}
    H=H_0+ \Delta \sum_{{\textbf{k}}} c_{{\textbf{k}}\uparrow}^{\dag}c_{-{\textbf{k}}\downarrow}^{\dag}+c_{-{\textbf{k}}\downarrow}c_{{\textbf{k}}\uparrow},
    \label{eq-h2}
\end{eqnarray}
where $\Delta$ represents the pairing strength. This model can be written in the Bogolibov-de Gennes (BdG) form,
$\mathcal{H}({\bf k})= \frac{1}{2}(d^x_{\textbf{k}}\sigma_0\otimes\sigma_x+ d^y_{\textbf{k}}\sigma_0\otimes\sigma_y+ d^z_{\textbf{k}}\sigma_z\otimes\sigma_z-\mu \sigma_z\otimes\sigma_0+
\Delta \sigma_y\otimes\sigma_y)$ under the Nambu basis $\psi_{\textbf{k}}^{\dag}=(c_{{\textbf{k}}\uparrow}^{\dag}, c_{{\textbf{k}}\downarrow}^{\dag},
c_{-{\textbf{k}}\uparrow},c_{-{\textbf{k}}\downarrow})$. The BdG Hamiltonian possesses time-reversal symmetry $\mathcal{T} \mathcal{H}^{\ast}({\textbf{k}}) \mathcal{T}^{-1} = \mathcal{H}(-{\textbf{k}})$
with $\mathcal{T}=\sigma_z\otimes \sigma_z$ and particle-hole symmetry $\Xi \mathcal{H}^{\ast}({\textbf{k}}) \Xi^{-1} = -\mathcal{H}(-{\textbf{k}})$ with $\Xi =\sigma_x\otimes\sigma_0$. Thus the
model (\ref{eq-h2}) belongs to two dimensional BDI class according to the ten-fold classification\cite{APS, kitaev}.
The combination of these two symmetries can give rise to a chiral symmetry $\mathcal{S}=\sigma_y\otimes\sigma_z$ with $\mathcal{S} \mathcal{H}({\bf k})\mathcal{S}^{-1}=-\mathcal{H}({\bf k})$,
which can bring us great convenience to analytically determine the topological phases and the associated phase boundaries.

\begin{figure}
    \includegraphics[width=0.45\textwidth]{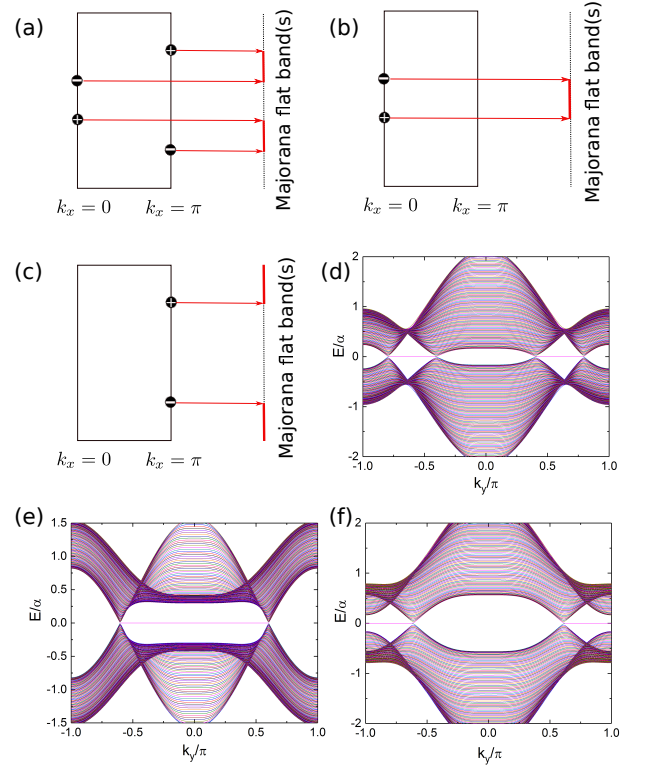}
    \caption{Edge states for a strip with width $L = 100$ along $k_y$ direction.
    (a), (b) and (c) show the Majorana flat bands in TG$^{\text{I}}(2, 2)$, TG$^{\text{I}}(2, 0)$
    and TG$^{\text{I}}(0, 2)$ phases, respectively. The corresponding
    spectra is shown in (d) - (f).}
    \label{fig-fig2}
\end{figure}

\begin{figure}
    \centering
    \includegraphics[width=0.45\textwidth]{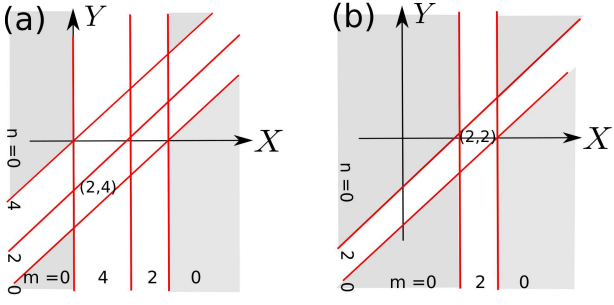}
    \caption{Phase diagrams in the presence of Zeeman field $h_z$ and $h_y$. (a) and (b) show the results with $|Z| < 1$
    and $|Z| > 1$, respectively. All topological TSCs are denoted by TG$^{\text{I}}(n, m)$ and the shadowed regimes mark
    the trivial phase, TG$^\text{I}(0, 0)$.}
    \label{fig-fig3}
\end{figure}

The gapped TSCs in two spatial dimension in BDI class is not permitted according to its classification\cite{APS, kitaev}, however, the nodal TSCs are still
allowed, which are characterized by winding numbers. Let $U\mathcal{S} U^\dagger = \text{diag}(1, -1)$, we find
\begin{equation}
    U \mathcal{H}({\bf k}) U^\dagger =
    \begin{pmatrix}
        0  & q_{\bf k} \\
        q^\dagger_{\bf k} & 0
    \end{pmatrix}, \quad
    U=\frac{1}{\sqrt{2}}\left(\begin{array}{cc}
\sigma_y &  \sigma_x \\
-\sigma_y & \sigma_x
\end{array}\right),
    \label{eq-q}
\end{equation}
with $q_{\bf k}=\mu \sigma_0+(d^x_{\textbf{k}}-i\Delta)\sigma_x-d^y_{\textbf{k}}\sigma_y +d^z_{\textbf{k}}\sigma_z$.
Thus $\text{det}(\mathcal{H}({\bf k})) = - \text{det}(q_{\bf k}) \cdot \text{det}(q_{\bf k}^\dagger)$ and the gap
closing condition is determined by $\text{det}(q_{\bf k})= \mu^2+\Delta^2-(d^{x}_{\textbf{k}})^2-(d^{y}_{\textbf{k}})^2-(d^{z}_{\textbf{k}})^2  + 2i  \Delta d^{x}_{\textbf{k}} = 0 $.
This equation shows that the gapless phases can only be realized at $k_x =0, \pi$, and the corresponding $k_y$ is determined by the following equations,
\begin{eqnarray}
(1-\cos{k_y})^2_{k_x = 0}= X, (1-\cos{k_y})^2_{k_x = \pi}= X-Y,
\label{eq-xy}
\end{eqnarray}
where the two independent parameters are
\begin{equation}
    X = {\mu^2+\Delta^2 \over \beta^2}, \quad  Y = {4\gamma^2 \over \beta^2}.
\label{eq-xy}
\end{equation}
In the vicinity of the nodal points, the effective Hamiltonian can be approximated as $h_{\text{eff}} = v_x \delta k_x \sigma_x + v_y \delta k_y \sigma_y$,
which is a type-I Dirac equation. These gapless phases throughout this work are denoted as TG$^{\text{I}}(n, m)$, where $n$ and $m$ corresponds to the
number of Dirac points at $k_x =0$, and $k_x = \pi$, respectively. The phase diagram as a function of these two independent parameters is
presented in Fig. \ref{fig-fig1}. For the model in Eq. \ref{eq-xy}, $n$ and $m$ may equal to $\{0, 2\}$ when $0 \le X \le 4$, or $0\le X-Y \le 4$. Thus we
have four different nodal TSCs, where TG$^{\text{I}}(0, 0)$ denotes the trivial gapped phase. It is necessary to emphasize that although the linear
SOC strength $\alpha$ do not explicitly enter the topological boundaries, it is essential for these gapless phases.

The emergence of these topological phases may be understood from the
effective pairings in the picture of dressed basis \cite{Chan}. Let
$\mathcal{H}_0 \psi_{\pm, {\bf k}} = \varepsilon_{\pm}({\bf k})
\psi_{\pm,{\bf k}}$, then we find $c_{{\bf k}\uparrow}^{\dag}
c_{-{\bf k}\downarrow}^{\dag}= \frac{d_{\bf k}^x+id_{\bf k}^y}{2
d_{\bf k}}[\psi^{\dag}_{-, {\bf k}}\psi^{\dag}_{-, -{\bf
k}}-\psi^{\dag}_{+, {\bf k}}\psi^{\dag}_{+, -{\bf
k}}+\sqrt{\frac{d_{\bf k}+d_{\bf k}^z}{d_{\bf k}-d_{\bf
k}^z}}\psi^{\dag}_{+, {\bf k}}\psi^{\dag}_{-, -{\bf k}}-
\sqrt{\frac{d_{\bf k}-d_{\bf k}^z}{d_{\bf k}+d_{\bf
k}^z}}\psi^{\dag}_{-, {\bf k}}\psi^{\dag}_{+, -{\bf k}}]$ with
$d_{\bf
k}=\sqrt{(d^{x}_{\textbf{k}})^2+(d^{y}_{\textbf{k}})^2+(d^{z}_{\textbf{k}})^2}$.
Thus all pairings, including inter-band and intra-band pairings, are
odd (even) functions of $k_x$ ($k_y$). These effective pairings are
resemblance to the unconventional pairings required for nodal TSCs
in previous literatures\cite{wenyu, shusa, ktlaw1, ktlaw2,
youjianbing, tanaka, sato, Kashiwaya, sato, Kashiwaya, RevSchnyder}.
The difference is that in this work these pairings can be controlled
in experiments by engineering the single particle Hamiltonian.

The robustness of these nodal TSCs can be understood from the winding number
\begin{eqnarray}
    N_1=\frac{1}{2\pi}\oint_{S^1} d{\bf k}\cdot
    \partial_{{\bf k}}\text{Im}[\ln\text{det}(q)],
    \label{eq-N1}
\end{eqnarray}
where the contour $S^1$ encloses only one of the nodal points in momentum space. In this case, $N_1 = \pm 1$ (see Fig. \ref{fig-fig2}). This number
can take other integer values when more than one nodal points are enclosed. The whole system is topologically neutral since these topological defects
should be created or destroyed in pairs with opposite winding numbers. In our model they are fused at $k_y = 0$ when $X = 0$, $Y=X$, and $k_y = \pi$ when
$X=4$, $Y=X-4$. These gapless phases may be a general feature in all TSCs, and in previous literatures these nodal phases were predicted based on
unconventional pairings.

The edge states in these nodal phases are different from these in the gapped TSCs. Fig. \ref{fig-fig2} plots the spectra of system for a strip along $k_y$ direction. We can find
dispersionless zero-energy flat bands for all these three different nodal TSCs. The observed edge states can be easily seized in a dimension reduction manner. In momentum space
considering $k_y$ as an external parameter, the system consists of a series of 1D subsystems paramtered by $k_y$. Each subsystem has a well-defined chiral symmetry $\mathcal{S}$
and is gapped as long as $k_y$ not cross the nodal points in momentum space. In this subspace, we consider the following topological invariant\cite{sato, ching, beri},
\begin{eqnarray}
w(k_y) =\frac{1}{2\pi} \int dk_x \partial k_x
\text{Im}[\ln\text{det}(q)].
    \label{eq-w}
\end{eqnarray}
The subsysem is topological nontrivial when $w(k_y) \ne 0$, which will give rise to zero energy edge state(s) when a boundary is imposed along $k_x$ direction.
For the phase in TG$^{\text{I}}(2, 2)$, we find $w(k_y = 0) = 0$, thus the two flat bands should be disjointed. In TG$^{\text{I}}(2, 0)$, we find
$w(k_y = 0) = -1$, thus the flat band should across the $k_y = 0$ point. This is different from the TG$^{\text{I}}(0, 2)$, where the flat band is connected through
the Brillouin zone at $k_y = \pi$ since $w(k_y = \pi) =1$. The similar analysis can be applied for strip along other directions. Note that the model is an
even function about $k_y$, these zero energy modes for each $k_y$ are connected by particle-hole symmetry, thus all the flat bands are essentially Majorana flat bands.

\begin{figure}
    \centering
    \includegraphics[width=0.48\textwidth]{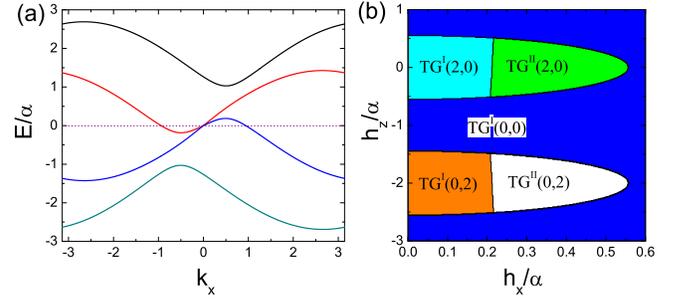}
    \caption{(a) Band structure for a typical type-II Dirac point; (b) Phase diagram
    for the two nodal points in the model with $h_x$ and $h_z$. Parameters are $\beta=1.2\alpha$,
$\gamma=\alpha$, $\Delta=0.2\alpha$, $\mu=0.6\alpha$,
$h_y=0.3\alpha$. In (a) $h_x=0.5\alpha$, $h_z=-0.1\alpha$,
$k_y=0.353$.}
    \label{fig-fig4}
\end{figure}

Next, we explore the fate of these nodal phases in the presence of external Zeeman fields by adding a term $H_z=\sum_{{\bf k}\sigma\sigma'}c_{{\bf
k}\sigma}^{\dag}{\bf h}\cdot \sigma c_{{\bf k}\sigma'}$ into model (\ref{eq-H0}). In the BdG formulism, $\mathcal{H}_z$ can be arranged into
$\mathcal{H}_z({\bf k}) = h_x\sigma_z\otimes\sigma_x+h_y\sigma_0\otimes\sigma_y+h_z\sigma_z\otimes\sigma_z$. The roles played by Zeeman fields
along different directions may be seized through its relation with chiral symmetry $\mathcal{S}$. The $h_y$ and $h_z$ terms anti-commute with
$\mathcal{S}$, whereas $h_x$ term commutes. Thus when $h_x=0$ the whole Hamiltonian still respects the chiral symmetry $S$ and the nodal points
can be found as
\begin{equation}
    (Z -\cos{k_y})_{k_x = 0}^2=X, (Z -\cos{k_y})_{k_x = \pi}^2= X-Y,
\end{equation}
where the three parameters are defined as,
\begin{equation}
    Z = 1 + {h_y \over \beta}, X = {\mu^2+\Delta^2-(2\gamma+h_z)^2 \over \beta^2}, Y = 4{\gamma^2 + \gamma h_z\over \beta^2}.
\end{equation}
The solution of the above equations depends strongly on the value of $Z$. If $|Z| > 1$, only two nodal points are allowed for each $k_x$; otherwise,
four nodal points are allowed. The corresponding phase diagrams as a function of $X$ and $Y$ for these two cases  are presented in Fig. \ref{fig-fig3}.
In the vicinity of the nodal point, the effective Hamiltonian can be described by type-I Dirac equation, thus all these phases are
still denoted by TG$^\text{I}(n,m)$.

\begin{figure}
    \includegraphics{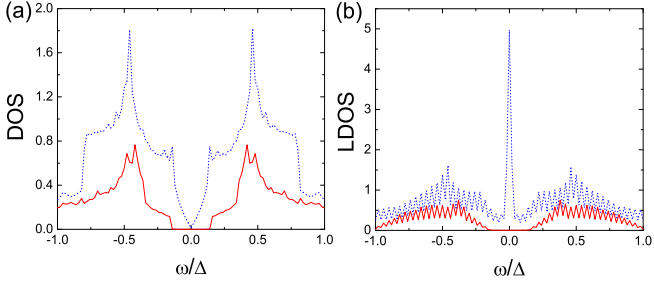}
    \caption{(a) The density of state (DOS) in the bulk and (b) the local density of state (LDOS) at the edge for gapped
    trivial phase (red solid line) and topological nodal phases (blue dashed line). The LDOS at zero energy is strongly
    enhanced by the Majorana flat bands. For the topological nodal phase, $\Delta=1.2\alpha$, $\mu=1.8\alpha$
    and for the gapped phase,  $\Delta=3.5\alpha$, $\mu=1.0\alpha$. Other parameters are:  $\beta=1.2\alpha$, $\gamma=\alpha$.
    The DOS in the bulk is determined by $\rho(\omega) = -\frac{1}{\pi}\sum_{{\bf k}}\text{Tr}[\text{Im}\mathcal{G}_{\text{r}}({\bf k},\omega)]$,
    where the retarded Green's function is defined as $\mathcal{G}_{\text{r}}({\bf k},\omega)=[\omega+i\delta-\mathcal{H}({\bf k})]^{-1}$. The LDOS
    at the edge is computed via the Green function iteration method in ref. \onlinecite{potterer}.}
\label{fig-fig5}
\end{figure}

The presence of $h_x$ can fundamentally change the symmetry of this model and reduce the system from BDI class to the D
class due to breaking of time-reversal symmetry. However the nodal points will not be immediately destroyed due
to the reason that the Hamiltonian is an even function of $k_y$, thus combining with the particle-hole symmetry
we always have at $k_x = 0$ or $k_x = \pi$
\begin{equation}
    \Xi \mathcal{H}^{\ast}({\bf k}) \Xi^{-1} = -\mathcal{H}(-k_x, k_y) = -\mathcal{H}({\bf k}).
    \label{eq-11}
\end{equation}
The first equality of above equation ensures a different $\mathbb{Z}_2$ invariant associated with these two
particle-hole invariant axes\cite{kitaev1, ghosh},
\begin{equation}
    \nu(k_y) = \text{sign}[\text{Pf}(W(0, k_y))\text{Pf}(W(\pi, k_y))],
\end{equation}
where $W(k_x, k_y) = \mathcal{H}\Xi$ and $W(k_x, k_y)^T = -W(k_x, k_y)$ when $k_x = 0, \pi$. We
find Pf$(W(0, k_y)) = h_x^2 -\Delta^2 - \mu^2 + \beta^2 (1
-\cos(k_y))^2$ and Pf$(W(\pi, k_y)) = h_x^2 + 4\gamma^2 -\Delta^2 -
\mu^2 + \beta^2 (1 -\cos(k_y))^2$. Thus for $k_x = 0$, the gapless point is determined by $(1-\cos{k_y})^2=X$; and for $k_x = \pi$, it
is determined by $(1 -\cos{k_y})^2=X - Y$, where $X = (\mu^2 +
\Delta^2 - h_x^2)/\beta^2$ and $Y = 4\gamma^2/\beta^2$. For this reason, the Majorana flat bands about these two special lines in the
Brillouin zone can still be found when $\nu(k_y) = - 1$.

The presence of $h_x$ will tilt the band structure along the $k_x$
direction, thus the effective Hamiltonian in the vicinity of the nodal points (${\bf k}_0$) can be written as
\begin{equation}
    h_\text{eff}({\bf k}_0 + \delta {\bf k}) = v_x \delta k_x \sigma_x + v_y \delta k_y \sigma_y + \epsilon \delta k_x,
\end{equation}
which may give rise to the type-II Dirac dispersion when $\epsilon^2
> v_x^2$. The presence of this tilting term will not change the chirality of
Dirac point, which can be determined as $\text{sign}(v_xv_y)$. The
boundary between the type-I and type-II Dirac points is thus
determined by $\epsilon^2 = v_x^2$. A typical type-II band structure
in our model is presented in Fig. \ref{fig-fig4}a, which is a gapless phase,
similar to that in Fulde-Ferrell-Larkin-Ovchinnikov superconductors\cite{Chan}.
In Fig. \ref{fig-fig4}b, we plot the phase diagram as a
function of Zeeman fields $h_z$ and $h_x$. We find that the type-I
topological nodal phase will be continuously driven to the type-II
nodal phase (denoted as TG$^{\text{II}}(n, m)$) when the
in-plane Zeeman field $h_x$ exceeds some critical value. In the
much stronger Zeeman fields the system will evolve to the trivial
gapped phase (TG$^\text{I}$(0,0)) through fusing of type-II Dirac points. More
intriguingly topological type-II nodal phases can be realized when
starting from other TG$^\text{I}(n, m)$ phases in Fig. \ref{fig-fig3},
which provides a general route to realize these intriguing phases.
In the TG$^\text{II}$ superconductors, the edge states may mix up with bulk bands, and the Majorana flat bands
may not be seen anymore. The Majorana flat bands in this new system can still be find in the TG$^\text{I}$
nodal phases, but now they are classified by $\mathbb{Z}_2$.

We finally address the random disorder effect, which is unavoidable in real materials. This effect can be taken into account via
$H_{\text{dis}} =\sum_{iss'}V_{iss'} c_{is}^{\dag}c_{is'}$, where $\overline{V_{iss'}V_{j\sigma\sigma'}} = V^2\delta_{i,j}\delta_{s\sigma}\delta_{s'\sigma'}$,
where $V$ is the disorder strength. For weak disorder and to the leading order, the disorder-averaged Green function can be calculated from the free
Green function $\mathcal{G}_0^{-1}(i\omega,{\bf k})=i\omega-H({\bf k})$ by $\overline{\mathcal{G}}(i\omega,{\bf
k})=[\mathcal{G}_0^{-1}(i\omega,{\bf k})-\Sigma(i\omega)]^{-1}$ with the self-energy $\Sigma(i\omega)=V^2\sigma_z\otimes\sigma_0\sum_{{\bf
k}}\mathcal{G}_0(i\omega,{\bf k})\sigma_z\otimes\sigma_0$ \cite{leepa, LiJian, Groth}. If the dependence of the self-energy on frequency can
be neglected under Born approximation, we find that the disorder can renormalizes all the momentum-independent parameters, such as chemical potential,
pairing strength and Zeeman fields. As we have demonstrated before, the gapless phases are also allowed in D class TSCs, thus the realized
nodal TSCs and the associated Majorana flat bands are robust against weak disorders. In experiments, these different phases can be
distinguished using density of state and local density of state measurement probed by scanning tunneling spectroscopy, which has proved to be an
effective tool to explore two-dimensional unconventional superconductors\cite{Kashiwaya, tanakaa, yada}; see simulation in the bulk and
the boundary in Fig. \ref{fig-fig5}.

To conclude, a route to realize nodal TSCs by direct engineering the single particle Hamiltonian, instead of pairings, is demonstrated.
We propose a minimal model based on linear and quadratic SOC for the realization of different nodal superconductors. In the vicinity of the nodal
points, the effective Hamiltonian can be either type-I or type-II Dirac Hamiltonians, and their crossover can be driven by the external Zeeman fields.
These nodal phases are robust against weak random perturbation, thus may be realized using realistic materials in proximity to
conventional $s$-wave superconductors. The Majorana flat bands protected by $\mathbb{Z}$ in BDI class or $\mathbb{Z}_2$
in D class superconductors can greatly enhance the local density of states along the boundaries, thus can serve as smoking-gun
evidence for experimental detection with scanning tunneling spectroscopy.

\textit{Acknowledgements.} B.H., X.Y. and N.X. are supported by
National Natural Science Foundation of China (No. 11547047, No. 11504143, No. 11404278).
M.G. is supported by the National Youth Thousand Talents Program (No. KJ2030000001), the USTC start-up funding (No. KY2030000053).

\end{document}